\documentclass{article}
\usepackage{spconf,amsmath}
\usepackage{CJK, hyperref, url}
\usepackage{amsmath,graphicx,bm}
\usepackage{amssymb,amsmath,epsfig, epstopdf, subfigure}
\usepackage{multirow}

\title{Bilingual Streaming ASR with Grapheme units and Auxiliary Monolingual Loss}

\name{Mohammad Soleymanpour, Mahmoud Al Ismail, Fahimeh Bahmaninezhad, Kshitiz Kumar, Jian Wu}
\address{Microsoft Speech and Language Group\\
{\small \tt m.soleymanpour@uky.edu, malismail@microsoft.com}}
\begin{document}

\ninept
\maketitle
\begin{abstract}
We introduce a bilingual solution to support English as secondary locale for most primary locales in hybrid automatic speech recognition (ASR) settings. Our key developments  constitute: (a) pronunciation lexicon with grapheme units instead of phone units, (b) a fully bilingual alignment model and subsequently bilingual streaming transformer model, (c) a parallel encoder structure with language identification (LID) loss, (d) parallel encoder with an auxiliary loss for monolingual projections. We conclude that in comparison to LID loss, our proposed auxiliary loss is superior in specializing the parallel encoders to respective monolingual locales, and that contributes to stronger bilingual learning. We evaluate our work on large scale training and test tasks for bilingual Spanish (ES) and bilingual Italian (IT) applications. Our bilingual models demonstrate strong English code-mixing capability. In particular, the bilingual IT model improves the word error rate (WER) for a code-mix IT task from 46.5\% to 13.8\%, while also achieving a close parity (9.6\%) with monolingual IT model (9.5\%) over IT tests.
\end{abstract}



\section{Introduction}\label{Sec:Introduction}
It's estimated that more than half of the world's population can speak multiple languages. It's therefore imperative to support code-mixing or fully bilingual scenarios for most locales \cite{heigold2013multilingual, liu2019multilingual}. English is a universal language so we specifically focus on building a strong English code-mixing attributes for other primary ASR locales. Besides bilingual scenarios, we also expect to improve the recognition of words borrowed from English to other locales.

Much of the recent progress in ASR has been pivoted towards End-to-End (E2E) models \cite{sak2017recurrent, chiu2018state,Bahdanau-AttentionASR,luscher2019rwth}. In comparison to hybrid models, E2E models are lexicon- as well as alignment-free, and have simpler training stages. However, E2E models are also data hungry and require significantly larger data to jointly learn the tightly coupled acoustic and language components. Hybrid models are composed of independent components in acoustic model, language model, pronunciation lexicon, and offer greater flexibility in training particular components over associated data resources. The flexibility manifests in greater adaptability for new scenarios with limited data. There has also been some work on borrowing learnings from hybrid model for E2E \cite{xieFactorizedTransducer} to improve the adaptability of E2E models.

While we expect to address E2E challenges in future iterations, we also see that hybrid ASR holds big applications in industrial settings. We have seen bilingual and broader multilingual treatment for E2E models \cite{pratap2020massively,dalmia2021transformer} but similar developments are missing for hybrid models. Consequently we dedicate this work to developing a code-mixing streaming ASR solution for hybrid model. Our key directions are: (a) transition from phone-units vocabulary to grapheme-units for hybrid models, (b) develop a fully bilingual solution with shared and parallel encoder networks, (c) develop LID loss or auxiliary monolingual projection loss to combine the parallel encoders and produce better learning. We explore and summarize key insights and evaluate our work on large scale tasks. We have 3 precise ways to monitor our progress: (1) minimal regression between bilingual and the corresponding monolingual locale over the primary task, (2) less than 10\% relative regression for bilingual and monolingual EN over EN task, (3) strong gains over English code-mixing tasks. The bilingual or multilingual work significantly increases the scope of training and our work too contributes to building robust acoustic embeddings for transfer learning low-resource locales \cite{huang2013cross, schultz2001experiments,  gales2015unicode, ghahremani2017investigation, TL_ASR_Wang,rabiee2023soft, giollo2020bootstrap, wang2021unispeech}.

Next we motivate a transition from phone units to grapheme-letter units for hybrid ASR in sec.~\ref{Sec:Motivation}. We describe our training stages, fully bilingual steaming Transformer model, and parallel encoder structure in sec.~\ref{sec:bilingual}. We present our experiments and results in sec.~\ref{Sec:Experiments} and conclude in sec.~\ref{Sec:Conclude}.

\section{Bilingual Hybrid ASR Lexicon Units }\label{Sec:Motivation}
The key components in hybrid ASR are lexicon, acoustic model, language model and decoder. Lexicon describes words as sequence of units, acoustic model produces a Softmax score over the clustered-contextual-units, and decoder produces the best hypotheses in the context of an external language model (LM). Acoustic phones form lexicon units for most large-scale  hybrid ASR systems. The phone units are determined by expert linguists and closely follow word pronunciations. These units are historically locale-dependent and are typically 30-45 units for each locale. A closer study of the hybrid acoustic units is critical to embed  code-mixing attributes for other locales. We naturally desire greater sharing among the locale-specific units for stronger bilingual learning. However, the phone units are created and optimized for particular locales that hinders across-lingual sharing. We describe limited sharing for phonetic lexicon for a common word ``president" in Table~\ref{table:PhoneticLex}, where, the phone ``eh" for IT corresponds to ``e" for ES but ``ax" for EN.  It's clearly non-trivial to learn phone mappings from one locale to another. There has also been work on International phone set (IPA) \cite{ladefoged1990revised} to unify and produce shared phone units but that too requires expert linguists to best map the historically built phone sets to IPA, and the mapping is typically lossy. 


{\small
\begin{table}
\begin{center}
\caption{{\it Locale-specific phonetic lexicon for a common word ``president".}}\label{table:PhoneticLex}
\begin{tabular}{|c|c|c|c|c|c|c|c|c|c|}
\hline
EN & p & r &  \bf{eh} & z & ih & d & \bf{ax} & n & t\\
\hline
IT & p & r &  e  & z & i  & d & \bf{eh} & n & t  \\
\hline
ES & p & r  & \bf{e}  & s & i  & d & \bf{e}  & n & t  \\
\hline
\end{tabular}
\end{center}
\end{table}
}

\subsection{Grapheme-letter units for bilingual ASR}\label{sec:units}
In this section we motivate grapheme units and specifically letter units for bilingual work. Grapheme-based ASR techniques have been extensively studied in \cite{kanthak2002context, eyben2009speech} and more recently in \cite{liu2019multilingual, wang2018phonetic, le2019senones,sak2017recurrent, chiu2018state}. Our current work targets Latin locales where grapheme units naturally exhibit strong across-lingual sharing. We develop a new hybrid lexicon where words are simply composed of letters instead of phones. These clustered-contextual-letters are also called ``chenones" \cite{le2019senones}. Following \cite{le2019senones} we also retain special units for the start and end of words, ex - ``president: \_p r e s i d e n \_t". We also introduce a simplification of our bilingual units by Romanizing accented variants like \{é, ê, ë\} to \{e\}. Our work finds connections with and extends grapheme systems in \cite{liu2019multilingual, le2019senones} by developing model structures and auxiliary loss functions to minimize accuracy gaps between monolingual and bilingual variants.

\section{Bilingual ASR - Shared+Parallel Encoders}\label{sec:bilingual}
The hybrid model training constitutes 3 key stages. We use Kaldi \cite{povey2011kaldi} to train a bilingual Gaussian Mixture Model-Hidden Markov Model (GMM-HMM) with grapheme-letter units described in sec.~\ref{sec:units}. Next we train a vanilla Time Delay Neural Network (TDNN) \cite{fathima2018tdnn}  to produce frame alignments and finally train a fully bilingual streaming Transformer \cite{lu2020exploring} model on large scale tasks. We discuss additional details, design considerations, training data, model parameters, and training criterion in sec.~\ref{Sec:Experiments}. Above bilingual model marks a significant milestone in this work. We also benchmark against monolingual models and identify gaps. We address the gaps by incorporating locale-specific learning in shared+parallel encoder architecture in the next sections. 

\subsection{Parallel Encoders}\label{sec:2encoders}
Bilingual acoustic models address our code-mixing objectives. However, a vanilla training with fully shared layers over the pooled monolingual corpora misses locale-specific learning opportunities. We explore that direction by augmenting our fully shared model structure with parallel encoder layers. Our design consideration is to minimize the total model size - we achieve that by sharing bottom model layers and incorporating a few parallel layers at the top. This design retains good scope for locale-specific improvements. It's also a natural extension of fully bilingual model as initial layers learn low-level common features, whereas, upper layers learn high-level features suitable for dedicated monolingual tasks. We coin the locale-specific layers as Parallel Encoders (PE).  Next we describe techniques to soft-combine the PE outputs to enable code-mixing scenarios.



\subsection{Parallel Encoders with LID loss}\label{sec:LID}
\begin{figure}[h]
\centering
{\includegraphics[width=0.35\textwidth]{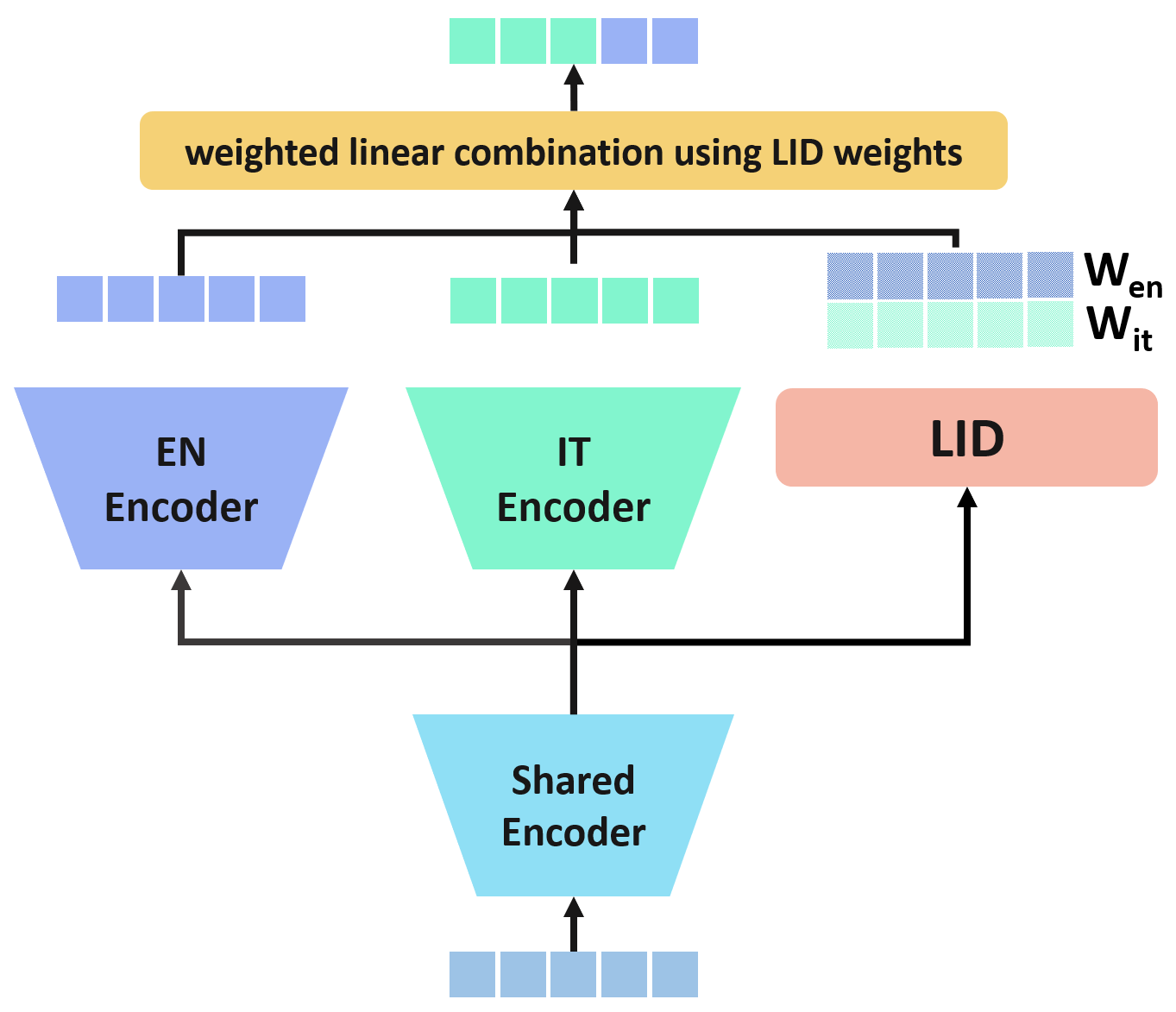}}
\caption{\it Parallel Encoders with LID. }
\label{Fig:parallel_encoder_lid}
\end{figure}

We explore a standard technique and develop an LID module to soft-combine the parallel encoders. The model architecture is noted in Fig.~\ref{Fig:parallel_encoder_lid}. The LID-based methods have been successful in multilingual E2E models \cite{dalmia2021transformer, jointAsrLidE2e}, as it dynamically weights embeddings from individual encoders. Therefore, we design LID to predict the probability of the locales, namely, \texttt{EN} and \texttt{IT}, as well as an additional class for ``silence", \texttt{sil} as \texttt{sil} is common to both locales. We use the probability of \texttt{EN} and \texttt{IT} to soft-combine the output of the PE while normalizing out the \texttt{sil} class. The LID is jointly optimized with the rest of the model using a cross-entropy LID criterion over the frames with a relative weight of $0.02$ for LID loss.
Jointly optimizing LID with the rest of the model presented some training stability issues. We chose to pre-train the core bilingual model (without LID module) using hard locale labels. There the shared layers are trained over data from both locales but PE layers are only be exposed to specific locales. Then we incorporate LID module and jointly finetune the PE and LID layers. This explicitly constrains the PE to specialize and enable monolingual scenarios. We also realize that an accurate LID is a difficult goal as many words are common for say EN and IT locales. This means that LID will invariably make mistakes during training as well as inference, and will promote some interdependence among the parallel heads, \emph{i.e.}, weaken orthogonality in the PE layers. We address aforementioned challenge by introducing locale-specific auxiliary losses for the PE structure in the next section.


\subsection{Parallel Encoders with auxiliary losses}\label{sec:3loss}
\begin{figure}[]
\centering
{\includegraphics[width=0.35\textwidth]{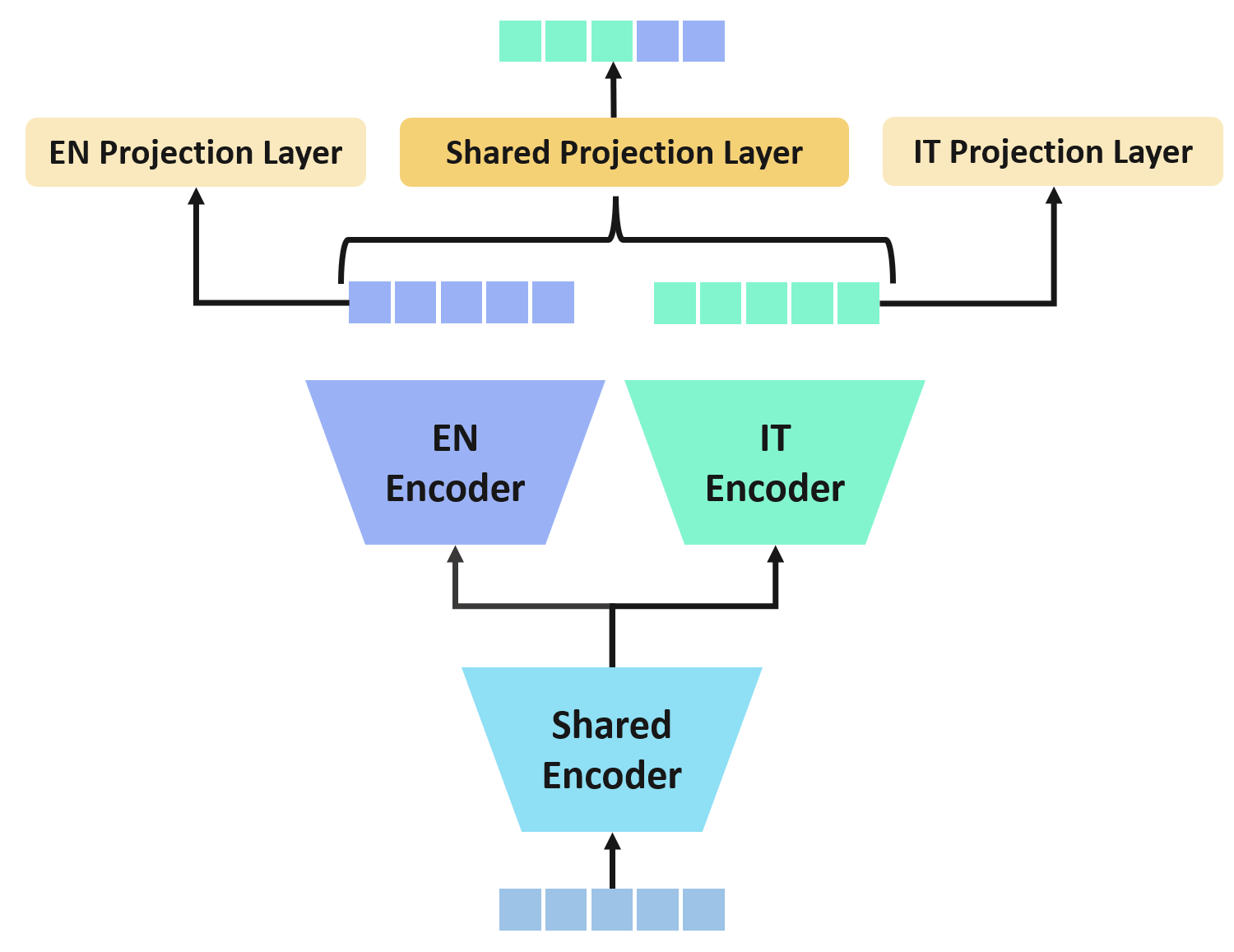}}
\caption{\it Parallel Encoders with auxiliary monolingual losses. }
\label{Fig:parallel_encoders_losses}
\end{figure}

In order to specialize the PE to specific locales we introduce monolingual auxiliary losses during training. This is done by projecting the locale specific hidden representations to independent feature spaces and constraining each encoder to predict locale-specific acoustic-states (\emph{chenones} or \emph{senones}). We adopt the idea of language-aware encoder \cite{langAwareEncoder} that includes an additional label assigned to locales foreign to the encoder. The hidden representations of the PE are concatenated and fed to a shared projection layer to learn a bilingual feature space.  This architecture is depicted in Fig. \ref{Fig:parallel_encoders_losses}. We use cross-entropy criterion for auxiliary losses that are jointly optimized with the rest of the model. We use the encoder-specific projection layers during training stage and simply use the shared projection layer during inference. We find that PE with auxiliary losses is more suitable solution as the model size and latency aren't that impacted during inference. 

\begin{table}
\begin{center}
\small
\caption{{\it Training and Test data.}}\label{table:Data}
\begin{tabular}{|c|c|c|c|c|c|}
\hline
Locales &  EN & IT & ES \\ \hline
Train Data [hrs] & 77k & 15k & 12.5k \\ \hline
Test Data  [utts] & 1.8M & 0.5M & 0.3M \\ \hline
\end{tabular}
\end{center}
\end{table}
    
\section{Experiments and results}\label{Sec:Experiments}
We note our training data in Table~\ref{table:Data}. The data is anonymized with personal identifiable information removed. The data reflects Microsoft usage scenarios including close-talk, far-field, natural conversation, short query, dictation, voice search, and call center tasks. We also apply a 3-fold data augmentation including noise, reverberation, and speed perturbation to ES and IT data. In our experiments, we report average WER on each locale over all the aforementioned tasks. In some cases, we present WER on some specific tasks (Agent, Conversations and Dictation) to highlight the effectiveness of the proposed techniques. The Code-Mixed task is simulated by joining shorter utterances from different locales. All the reported experiments use a 5-gram LMs with over 1M vocabularies. For bilingual work we conduct interpolation of monolingual n-grams LMs with a bias towards the primary locale. Our ASR features are 160-dim log-Mel values, we refer to  \cite{kskumar_Bandpass_160d} for  some setup details. As previously mentioned, we build a DNN-HMM system using the corresponding streaming Transformer-based acoustic model with cross-entropy (CE) \cite{hinton2012deep} distributed training criterion for ASR. Our streaming Transformer model is 300 MB that consists of 18 layers with a block size of 624, multihead attention with 8 heads, and a feed-forward hidden size of 2048. The acoustic model is converted to ONNX format and compressed prior to evaluation over a batch size of 28 frames. We use ADAM optimizer \cite{kingma2014adam} for training Transformer models.



\subsection{Bilingual GMM-HMM and TDNN alignment model}

Here we describe our bilingual GMM-HMM and TDNN models. We use Kaldi \cite{povey2011kaldi} to train GMM-HMM and internal Microsoft tools to train other models. The objective of GMM-HMM is to produce clustered-contextual states, and that for TDNN is to produce good quality frame alignments. In our prior monolingual work we found it effective to train GMM-HMM and TDNN alignment models over a smaller scale 5k hrs of training data. We borrow that design for bilingual work as well except that we select 5k hrs for the primary locale (ex - ES or IT) and 3k hrs of EN data to retain a bias for the primary locale. We train GMM-HMM models with standard 80-dimensional mel-frequency cepstral coefficients (MFCC). The number of \emph{senones} (phone units) or \emph{chenones} (grapheme-letter units) are  $\sim8k$.

We benchmark bilingual TDNNs against respective monolingual TDNN models in Table~\ref{table:IT-ES-EN-TDNN}. The bilingual IT is within 7\% word error rate relative (WERR) of the monolingual IT (26.1\% vs. 24.4\%). Similarly, the bilingual IT+EN evaluated on EN is within 9.6\% WERR (26.1\% vs. 23.6\%) of monolingual EN and meets our stated goals in sec.~\ref{Sec:Introduction}. In Table~\ref{table:IT-ES-EN-TDNN} we see relatively larger WERR gap for bilingual ES+EN TDNN vs. monolingual ES. We realize that TDNN models are primarily to produce good alignments as we eventually train Transformer models with full scale data noted in Table~\ref{table:Data}. We can best evaluate TDNN alignments in the context of Transformer model results in Table~\ref{table:IT_WERs} and Table~\ref{table:ES_WERs}, and retrospectively find the bilingual TDNN alignments suitable for our bilingual work.


\begin{table}
\begin{center}
\small
\caption{{\it WERs [\%] for monolingual and bilingual TDNNs.}}\label{table:IT-ES-EN-TDNN}
\begin{tabular}{|l|c|c|c|c|c|}
\hline
Models &  Mono   &  Mono  & Mono  & Bilingual   & Bilingual\\ 
   &     ES     &  IT     & EN    &   ES+EN     & IT+EN    \\ \hline
\textbf{ES} &  18.9      & -       & -     &   22.5      & -         \\ \hline
\textbf{IT} & -          & 24.4    & -        & -        & 26.1     \\ \hline
\textbf{EN} & -          & -       & 23.6     & 25.8     & 26.1     \\ \hline
\end{tabular}
\end{center}
\end{table}

\subsection{Monolingual vs. Bilingual Transformer models}
We build on the TDNN alignments and train streaming Transformer \cite{lu2020exploring} models over large scale data in Table~\ref{table:Data}. We report WERs for bilingual IT+EN and bilingual ES+EN in Table~\ref{table:IT_WERs} and Table~\ref{table:ES_WERs}, respectively. There, we train 3 models for each locale (1) baseline monolingual model with standard phone units, (2) bilingual model from bilingual alignments over grapheme units descrbied in sec.~\ref{sec:units}, (3) train model\#2 on just the primary locale data \emph{i.e.} train monolingual model from bilingual alignments. These 3 models help us draw interesting conclusions. We report that bilingual IT shows superior results over the monolingual IT baseline (9.82\% vs. 10.1\%) on IT task, and yet achieves a strong bilingual performance on EN task (11.32\%). As we later report in Table~\ref{table:IT-EN-WER}, above bilingual EN WER of 11.32\% is close to the upper limit of 10.95\% from monolingual EN. We derive similar conclusions for bilingual ES+EN model in Table~\ref{table:ES_WERs}, where the bilingual ES+EN model shows 3\% WERR over the monolingual ES baseline (8.52\% vs. 8.78\%), along with 11.13\% WER on EN task. We also show an opportunity to train monolingual models over the bilingual grapheme alignment. This new recipe shows 5.7\% WERR for IT (9.52\% vs. 10.1\%) along with 4.4\% WERR for ES (8.39\% vs. 8.78\%). We also conclude that bilingual alignment over grapheme units are more suitable for knowledge sharing across locales than traditional phone units. Table~\ref{table:IT-EN-WER} shows a continuation of previous bilingual IT results in Table~\ref{table:IT_WERs}. There we report WERs for bilingual IT+EN along with corresponding monolingual IT, and monolingual EN models. All of above models are trained over alignments  from bilingual TDNN model over grapheme units. We note that the bilingual system shows 71\% WERR (from  46.3\% to 13.4\%) for the code-mixed IT task while also achieving a close parity (9.8\%) with monolingual IT model (9.5\%) over IT tests. We utilize these monolingual model results as the upper limit of expected performance from bilingual work. We also set them as  baselines for subsequent sections to guide PE developments.

\begin{table}
\centering
\caption{{\it WERs [\%] for monolingual and bilingual IT.}}\label{table:IT_WERs}
\begin{tabular}{|l|c|c|c|c|} 
\hline
Model                        & Mono           & Bilingual (1)   & Train (1) on IT       \\ 
Units						 & [phone]  & [grapheme]  & [grapheme]         \\ \hline
\textbf{IT}                  & 10.1           & 9.82            & 9.52                          \\ \hline
\textbf{EN}                  & -              & 11.32           & -                         \\ \hline
\end{tabular}
\end{table}

\begin{table}
\centering
\caption{{\it WERs [\%] for monolingual and bilingual ES.}}\label{table:ES_WERs}
\begin{tabular}{|l|c|c|c|c|} 
\hline
Model                        & Mono       & Bilingual (1)  & Train (1) on ES\\ 
Units						  &   [phone]  &  [grapheme]  & [grapheme] \\ \hline
\textbf{ES}                   & 8.78      & 8.52     & 8.39        \\ \hline
\textbf{EN}                   & -         & 11.13    & -       \\ \hline
\end{tabular}
\end{table}


\begin{table}
\begin{center}
\small
\caption{{\it WERs [\%] for bilingual IT+EN and equivalent monolingual using graphemes with bilingual alignments.}}\label{table:IT-EN-WER}
\begin{tabular}{|l|c|c|c|c|c|}
\hline
Models              & Bilingual (1) &  Mono IT          & Mono EN    \\ 
                   & IT+EN          & Train (1) on IT   & Train (1) on EN    \\ \hline
\textbf{IT}         & 9.82 & 9.52       & 71.39                \\ \hline
\textbf{EN}         & 11.32 & 77.71      & 10.95                 \\ \hline
\textbf{Code-Mixed} & 13.38 & 46.32      & 37.11                 \\ \hline
\end{tabular}
\end{center}
\end{table}

\subsection{Parallel Encoders} 
In this section, we present our experiments on PE with LID and PE with auxiliary losses described in sec.~\ref{sec:LID} and sec.~\ref{sec:3loss}, respectively. Our bilingual model consists of 18 Transformer layers. For PE work, we consider 15 shared layers and 3 layers per parallel encoder. Our LID module is comprised of 2 small Transformer layers. This PE with LID increases the model size from 300 MB to 400 MB. In Table \ref{table:pe_branch_wer}, we evaluate the locale-specific encoders within PE model prior to fine-tuning with LID. The IT-encoder branch shows 2\% WERR over bilingual model (9.61\% vs. 9.82\%) on IT tasks. This shows that  introducing few locale-specific layers is effective in finetuing the model to monolingual tasks. The code-mixed data has higher WER  for both the IT and EN encoders; that's expected as PE layers encourage monolingual attributes.

In Table \ref{table:pe_aux_1loss_wer}, we present WER for PE with LID and PE with/without auxiliary losses. We observe that PE with LID doesn't even improve over the purely bilingual model. This implies that LID-based soft-combination is ineffective in: (1) maintaining any locale-specific encoder specializations, and, (2) combining locale-specific encoders to support code-mixed tasks. Our LID has respectively 90\% and 85\% accuracy on EN and IT tasks. We also observe that PE with auxiliary loss is far more superior than PE with LID and bilingual model. This improvement likely stems from the model's ability to maintain encoder specialization while simultaneously combining locales with the shared projection layer. We notice that the PE with auxiliary loss rectifies mistakes over phrases with similar pronunciations such as ``di" in IT and ``the" in EN. 

As an important study, we also provide PE model without monolingual auxiliary loss in Table~\ref{table:pe_aux_1loss_wer}. There we simply train the shared projection layer. Comparing 9.68\% w/Aux vs. 10.04\% w/o Aux loss,  we conclude that auxiliary losses are clearly advantageous in guiding PE towards locale specialization and subsequently towards better bilingual learning. We noticed a small regression on the code-mixed task from PE w/Aux loss (13.32\%) over PE w/o Aux loss (12.89\%) - we hypothesize that this is tied to  simple shared projection layer in current implementation and can be improved by advanced mechanism (\emph{e.g} attention blocks). We leave this investigation for a future work. As an ablation study we also evaluated the IT projection layer in the bilingual PE w/Aux loss and report that IT WER as 9.5\%. That achieves our best reported monolingual IT results with bilingual grapheme units in Table~\ref{table:IT-EN-WER}. This has significant implications as training w/Aux loss can deliver both monolingual and bilingual models in a single-shot training. This  simplifies model deployment and serving as we can simply switch the projection layer to meet monolingual vs. bilingual objectives. Most shared layers with a few locale-specific layers are also critical for multilingual applications in resource-constrained settings. In Table~\ref{table:it_agent_recos} we provide some examples to show clear improvements from bilingual model on words borrowed from English. In a future work, we will extend our work to non-Latin locales, and also apply learnings to E2E models.

\begin{table}
\centering
\caption{{\it WERs [\%] for locale-specific encoders within PE model prior to fine-tuning with LID.}}\label{table:pe_branch_wer}
\begin{tabular}{|l|c|c|c|c|} 
\hline
Locales                        & IT Encoder & EN Encoder  \\ \hline
\textbf{IT}                     & 9.61       & 45.57       \\ \hline
\textbf{EN}                         & 46.77      & 11.26       \\ \hline
\textbf{Code-Mixed}                 & 28.98      & 24.6        \\ \hline
\end{tabular}
\end{table}

\begin{table}
\centering
\caption{{\it WERs [\%] of PE using the two techniques: LID, and with/out auxiliary losses (Aux.)}}\label{table:pe_aux_1loss_wer}
\begin{tabular}{|l|c|c|c|c|} 
\hline
 Locales                      & Bi-        & PE + LID     & PE & PE \\  
                       &  lingual         &              &  w/o Aux.          & w/ Aux.          \\  \hline
 
\textbf{IT} (All)        & 9.82             & 9.83         &  10.04               & \textbf{9.68}           \\  \hline
\hspace{3mm} Agent            & \textbf{5.3}     & 5.1          & 5.6                  & 5.5                     \\  \hline
\hspace{3mm} Conversation     & 8.7              & 8.8          & 8.9                  & \textbf{8.6}            \\  \hline
\hspace{3mm} Dictation        & 5.7              & 5.9          & 6.0                  & \textbf{5.7}            \\  \hline
\textbf{EN}                   & \textbf{11.32}   & 11.4        & 11.4                 & 11.45                   \\  \hline
\textbf{Code-Mixed}           & 13.38            & 18.3         & \textbf{12.89}       & 13.32                   \\  \hline
\end{tabular}
\end{table}

\begin{table}
\centering
\caption{{\it Recognition examples from Agent IT task between monolingual and bilingual models. The errors are underlined.}}\label{table:it_agent_recos}
\begin{tabular}{|ll|} 
\hline
\it Tran.                   & cerca today's special value~ su ~ qvc                                                               \\ 
\it Mono                    & cerca today~~~ special \underline{vallio} su \underline{q} \underline{sei}                          \\ 
\it Bili                    & cerca today's special value~ su ~ qvc                                                               \\ \hline
\it Tran.                   & morgana formatta ~scheda ~sd                                                                 \\ 
\it Mono                    & \underline{organa} ~~~\underline{formate}~~ scheda~ sd                                       \\ 
\it Bili                    & morgana formatta~ scheda~ sd                                                                         \\ \hline
\it Trans.                  & avvia microsoft edge                                                                                \\ 
\it Mono                    & avvia microsoft \underline{~~~~~~~~}                                                                \\ 
\it Bili                    & avvia microsoft edge                                                                                \\ \hline
\end{tabular}
\end{table}



\section{Conclusion}\label{Sec:Conclude}
We identified limited progress in bilingual solutions for hybrid ASR and guided this work to address those gaps. We formulated our work to support English as secondary locale for primary locales in the Latin language family. We motivated a transition from traditional phone units lexicon to grapheme units and more specifically to letter units. We developed a fully bilingual model and subsequently a model with parallel encoder layers. We studied and presented methods to combine the parallel encoders, where we showed that including auxiliary monolingual projection loss leads to effective bilingual learning. We evaluated our work on large scale bilingual ES and IT tasks, and achieved strong English code-mixing attributes (71\% WERR). Our bilingual work achieved 3-4\% WERR over the monolingual phonetic baseline, and still produced very competitive results on EN task. 

\vfill\pagebreak

\bibliographystyle{IEEEtran}
\bibliography{strings,refs}

\end{document}